\documentclass{aastex}
\newcommand{\pdt}[1]{{{\partial #1}\over {\partial t}}}
\newcommand{\kapp}{\kappa_P}
\newcommand{\kapr}{\kappa_R}
\newcommand{\er}{E_{R}}
\newcommand{\bbody}{B\left(T\right)}

\begin{document}
\title{The Impact of Circumplantary Jets on Transit Spectra and Timing
  Offsets for Hot-Jupiters}

\author{Ian Dobbs-Dixon$^{1,2,3}$,  Eric Agol$^{1,3,4}$, and Adam Burrows$^{3,5}$}

\affil{$^1$Department of Astronomy, Box 351580, University of Washington, Seattle, WA 98195}
\affil{$^2$Carl Sagan Postdoctoral Fellow; ianmdd@gmail.com}
\affil{$^3$Kavli Institute for Theoretical Physics, Santa Barbara, CA}
\affil{$^4$Miller Visiting Professor, Department of Astronomy, University of California, Berkeley, CA 94720-3411}
\affil{$^5$Department of Astrophysical Sciences, Peyton Hall, Princeton University, Princeton, NJ 08544}

\keywords{hot-Jupiters, atmospheric dynamics, radiative transfer, spectra, HD209458b} 
\begin{abstract}
We present theoretical wavelength-dependent transit light curves for
the giant planet HD209458b based on a number of state of the art 3D
radiative hydrodynamical models. By varying the kinematic viscosity in
the model we calculate observable signatures associated with the
emergence of a super-rotating circumplanetary jet that strengthens
with decreased viscosity. We obtain excellent agreement between our
mid-transit transit spectra and existing data from Hubble and Spitzer,
finding the best fit for intermediate values of viscosity.  We further
exploit dynamically driven differences between eastern and western
hemispheres to extract the spectral signal imparted by a
circumplanetary jet. We predict that: (i) the transit depth should
decrease as the jet becomes stronger; (ii) the measured transit times
should show timing offsets of up to 6 seconds at wavelengths with
higher opacity, which increases with jet strength; (iii)
wavelength-dependent differences between ingress and egress spectra
increase with jet strength; (iv) the color-dependent transit shape
should vary more strongly with wavelength for stronger jets. These
techniques and trends should be valid for other hot Jupiters as
well. Observations of transit timing offsets may be accessible with
current instrumentation, though the other predictions may require the
capabilities JWST and other future missions. Hydrodynamical models
utilized solve the 3D Navier-Stokes equations together with decoupled
thermal and radiative energy equations and wavelength dependent
stellar heating.
\end{abstract}

\section{Introduction}
Recent decades have witnessed an explosion of observed extrasolar
planets with a wide range of physical properties. Amongst these
planets, a subset transit their host star and allow for a number of
new observational techniques aimed at determining interior and
atmospheric composition, dayside temperature, and efficiency of energy
redistribution. In addition to primary transit measurements
\citep[eg]{charbonneau2000} these tools include secondary eclipse
\citep[eg]{deming2005}, differential spectroscopy
\citep[eg]{charbonneau2002}, phase-curve monitoring
\citep[eg]{harrington2006,cowan2007}, Doppler absorption spectroscopy
\citep{snellen2010}, and transit spectra \citep[eg]{brown2002}.

In hopes of understanding the nature of the atmospheric dynamics and
its role in both redistributing incident stellar energy throughout the
atmosphere and (potentially) influencing the overall evolution of the
interior, multiple groups are working on coupled
radiation-hydrodynamical solutions. Multiple assumptions and
approaches have been taken to both the radiative and dynamical
components, including relaxation methods ({\it i.e.}  Newtonian
heating) \citep{showman2002, showman2008_2, cooper2005, cooper2006,
  langton2007, langton2008, menou2009, rauscher2010}, kinematic
constraints designed to represent incident flux \citep{cho2003,
  cho2008, rauscher2008}, 3D flux-limited diffusion with
\citep{dobbsdixon2010} and without
\citep{burkert2005,dobbsdixon2008_1} a separate radiation component,
1D two-stream approximation \citep{heng2011}, and 1D
frequency-dependent radial radiative transfer
\citep{showman2009}. Approaches to the dynamical portion of the model
have included solving the equivalent barotropic equations
\citep{cho2003, cho2008, rauscher2008}, the shallow water equations
\citep{langton2007, langton2008}, the primitive equations
\citep{showman2002, showman2008_2, showman2009, cooper2005,
  cooper2006, menou2009, rauscher2010, heng2011}, Euler's equations
\citep{burkert2005,dobbsdixon2008_1}, and the Navier-Stokes equations
\citep{dobbsdixon2010}.

Multiple approaches to solving this difficult problem are quite
valuable. However, as observational capabilities improve we would like
to begin to distinguish between different numerical approaches and
parameter choices. The day-night temperature contrast from primary
transit and secondary eclipse measurements and phase dependent
measurements have proven useful in this respect. Many simulations
listed above produce strong circumplanetary eastward equatorial jets
which predict that the hottest point will be advected downwind. An
analytic model for the formation of these jets by \citet{showman2011}
suggests that the planetary-scale temperature differential excites
eddies associated with Rossby and Kelvin waves which lead to the
transfer of zonal angular momentum toward the equator. Calculations of
eddy angular momentum transport within the simulations presented here
seem to corroborate their theory. Observations of HD189733b
\citep{knutson2007, knutson2009, agol2010} appear to have convincingly
demonstrated the existence of equatorial super-rotation, showing that
the hottest point in the phase curve appears slightly before secondary
eclipse, as first predicted by \citep{showman2002}.  However, the
$3.5$-hour ($24^{\circ}$) offset of the phase variation observed at 8
microns is smaller than that predicted by General Circulation Models
(GCM) with solar abundance and synchronously rotating interiors
\citep{showman2009}, which may indicate that the atmospheric
velocities may have been overestimated. As GCMs are incompressible, it
is possible that the velocity, which is supersonic, may have been
overestimated resulting in a larger phase offset than observed.
Observations of $\upsilon$-Andromeda b \citep{crossfield2010} indicate a much
larger phase shift of $89^{\circ}$, completely out of the range of
current model predictions. This large phase shift also appears at odds
with the large amplitude of the phase curve (which has decreased and
shifted by 90 degrees with a re-observation and re-analysis of the
original data); however, the fact that this planet does not transit
its host star makes it a difficult case to decipher due to the unknown
mass, radius, and orbital inclination of the planet. Suffice it to
say, more work is necessary before we can assemble a coherent picture
of these objects.

Unfortunately, much of the detailed spatial and temporal structure
that is seen in numerical simulations is hidden in the necessarily
hemispherically averaged phase curves \citep{cowan2008}.  The phase
curve can be translated into the longitudinal brightness distribution
on the planet, but higher frequencies are strongly suppressed, so only
coarse features may be resolved \citep{cowan2008}.  Temperature
differences across jets, latitudinal dependence, vortices, and other
interesting sub-hemisphere scale phenomena largely disappear. However,
one technique that may prove quite useful in this respect is transit
spectroscopy. Taken as the planet transits its host star, transit
spectroscopy measures the absorption of stellar light by the upper
limbs of the planetary atmosphere yielding a wavelength-dependent
radius for the planet \citep{seager2000}. Given the viewing geometry
of the star, planet, and observer, such a measurement probes the
meridians delineating the day and night hemispheres (the terminators).
The variation in opacity with wavelength can cause the planet to vary
in absorption radius by $\sim 5 h$ \citep{burrows2003}, where $h$ is
the atmospheric scale-height, leading to depth variations on the order
of $10R_ph/R_*^2 \approx$0.1\% for $5h \approx 3500$ km, $R_p \approx
R_J$, and $R_* \approx R_\odot$.

Dynamics plays a crucial role in shaping the temperatures across the
terminators both at the surface and at depth. It is here that one
expects the largest deviations from radiative equilibrium. As
high-velocity jets advect energy across the terminator to the
nightside, the flow will cool radiatively, thus one would expect the
largest nightside temperatures to be closest to the terminator. Indeed
simulations show similar behavior, but the exact temperature
distribution depends sensitively on the details of the flow structure
and radiative and advective efficiencies. Transit spectra taken
both at mid-transit and during ingress and egress have great potential
to help distinguish between models and adopted parameters. Several
models presented in \citet{dobbsdixon2010} show pronounced variability,
with the largest amplitudes at the terminator region. Targeting the
terminator regions with multiple transit spectra measurements
may reveal spectral changes due to such dynamically driven weather.

Several other groups have also explored the differences between
transit spectra calculated with 1D or 3D
models. \citet{fortney2010} and \citet{burrows2010} perform similar
calculations utilizing the 3D GCM models of \citet{showman2009} and
\citet{rauscher2010}, respectively. As mentioned above, the dynamical
and radiative methods differ amongst these models and ours, but the
method for calculating the resulting spectra from the results is
largely equivalent.

In this paper we utilize the 3D pressure-temperature profiles
calculated using the 3D Navier-Stokes equations coupled to
wavelength-dependent stellar heating and 3D flux-limited diffusion for
the re-radiated component. Models differ from \citet{dobbsdixon2010}
in several respects; we now allow for advective flow over the polar
regions, stellar heating is modified to be both wavelength dependent
and to account for the slant optical depth of stellar light, and the
upper boundary condition for the radiation is improved. We describe
these modifications in Section 2 along with the method for calculating
transit spectra. In Section 3 we illustrate variations in the
predicted transit spectra of models with varying viscosity. In
essence, this is a proxy for the presence or absence of a
super-rotating equatorial jet discussed above. Because the variations
at mid-transit are somewhat small we also propose three methods for
extracting the signal utilizing the temporally dependent transit
spectra. Finally, we compare our models to existing data in the
literature. We conclude in Section 4 with a discussion of the results
and future detectability of such effects with the next generation of
instruments.

\section{Modeling Methodologies}
Calculating theoretical transit spectra has been done in a number
of ways, the most common of which involves a 1D solution coupling a
radiative transfer routine to the assumption of vertical hydrostatic
equilibrium. Here we move away from assumptions of both spherical
symmetry and hydrostatic equilibrium by utilizing 3D radiation
hydrodynamic models coupled together with post-processing calculations
to determine the detailed wavelength dependent spectra. This allows us
to produce potentially observable metrics by which we may constrain
the initial 3D models. We first discuss the hydrodynamical models
(particularly recent modifications to the stellar energy deposition,
flow in the polar regions, and flux-limited diffusion) followed by the
methods used to calculate the spectra.

\subsection{3D Dynamical Modeling}
To calculate the pressure and temperature throughout the atmosphere of
HD209458b we utilize a fully non-linear, coupled radiative
hydrodynamical code. We solve the fully compressible Navier-Stokes
equations throughout the 3D atmosphere together with coupled thermal
and radiation energy equations. Direct stellar heating of the planet
is calculated using a fully wavelength-dependent procedure. Local 3D
radiative transfer and cooling is included through multi-temperature
flux-limited diffusion. This model can self-consistently produce both
the radiative and dynamically induced inversions, as suggested in the
observed daytime spectra \citep{knutson2008}.

This model has been utilized to study a range of parameters, most
notably the viscosity \citep{dobbsdixon2010} and opacity
\citep{dobbsdixon2008_1}. Imposed viscosity, beyond that which is always
present in numerical simulations, is meant to represent explicit
dissipation processes that may not be captured in our numerical models
due to limited resolution or neglected physics. More importantly for
our present discussion, it allows us to study a range of flow
structures and determine how one might distinguish between them
observationally.

We have implemented several important changes to the numerical model
of \citet{dobbsdixon2010}: allowing fluid to flow over the polar
regions, altering the form of the energy deposition term in the
thermal energy equation, and altering the flux-limited diffusion
boundary conditions. We discuss each of these changes
below. Additional details regarding the hydrodynamical method can be
found in \citet{dobbsdixon2010}.

All simulations are run with a radial, longitudinal and latitudinal
resolutions of $\left(n_r,n_{\phi},n_{\theta}\right) =
\left(200,160,64\right)$. These correspond to individual grid cells of
$\left(\delta r,\delta\phi,\delta\theta\right) = \left(90 km,
2.25^{\circ}, 2.8^{\circ}\right)$. Comparing to the typical pressure
scale-height in Table (\ref{table:one}) there are more then $5$ radial
cells per scale-height. Typical timesteps are quite small, averaging
around $10$ seconds, primarily set by radial velocity. The pressure in
the simulation ranges from $10^4$ to $10^{-5}$ bars. Note this is much
deeper then any previous simulations and includes a fully convective
interior below the radiative zone traditionally simulated in
irradiated planets. The important effects of convection will be
discussed elsewhere.

\subsubsection{Polar-Traversing Flow}
In the models of \citet{dobbsdixon2010}, the choice of a spherical
grid, coupled with finite computational resources, necessitated
implementing a meridional boundary condition at high latitudes. In
spherical coordinates the size of the longitudinal grid cell decreases
rapidly at high latitudes. For codes requiring satisfaction of the
Courant-Levy-Fredreich (CFL) condition \citep{courant1928} a shrinking
grid size will drive the computational timestep to very low values,
grinding computation to a halt. Finding a latitude that allowed
significantly unencumbered flow, while not halting computation
motivated the choices in \citet{dobbsdixon2010}.

\begin{figure}
\plotone{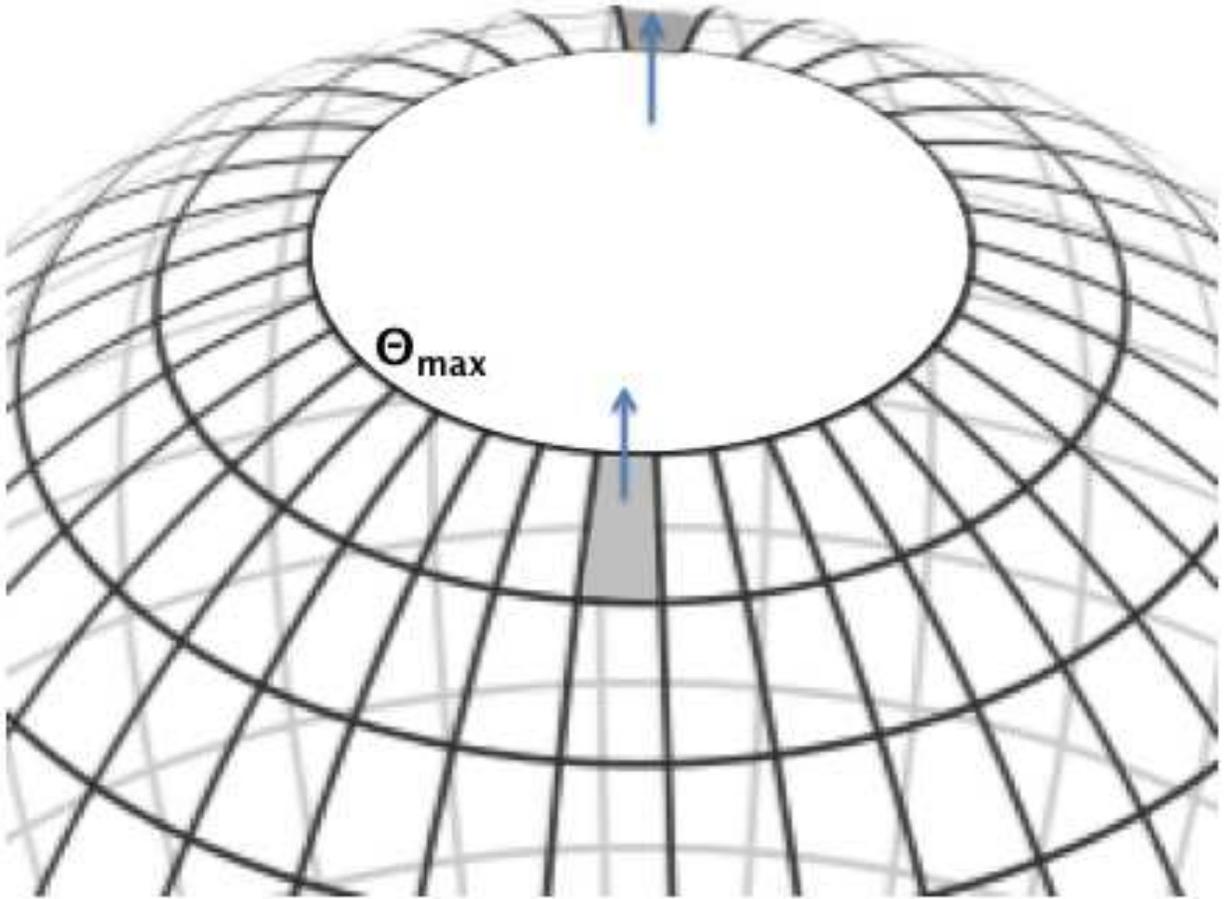}
\caption{An illustration of the staggered grid setup near the north
  pole. Scalars are defined at grid centers while velocities and
  fluxes are defined along appropriate grid edges. Cells along the
  polar circle
  ($\left(\phi,\theta\right)=\left(\phi,\theta_{max}\right)$) utilize
  the cell at $\left(\phi+\pi,\theta_{max}\right)$ as a neighboring
  cell. For instance, the two shaded cells communicate directly and
  the velocities (blue arrows) across the cell edges are identical.}
\label{fig:pole}
\end{figure}

In the results presented here we attempt to relax this condition with
an approximation that allows meridional flow over the poles. Figure
(\ref{fig:pole}) shows a schematic of the computational region around
the northpole region. We utilize a staggered grid, with velocities
calculated at the appropriate cell edges and scalars defined in the
center \citep{kley1987}. In our scheme, whatever flows northward from
one cell immediately flows into the cell on the opposite side of the
pole, this includes all meridional fluxes required for the advection
scheme. In essence, each cell along the $\pm\theta_{max}$-line behaves
as though the cell at $\phi=\phi+\pi$ is adjacent to it. One can
choose whatever value of $\theta_{max}$ you would like. In the limit
that $\theta_{max}=\frac{\pi}{2}$ the scheme is completely
accurate. In deference to the CFL limitations, we take the meridional
size of the polar region ($2\theta_{max}$) to be slightly larger than
the $\delta\theta$ of the rest of the grid so that we may take
reasonable time steps. Strictly speaking, this violates causality, as
meridional flow is able to transit the entire polar region within a
single timestep.

There are three reasons we believe our treatment of the pole may be
justified. The first is that, to first order, we expect the model to
relax to an approximately steady-state solution characterized by a
large dayside driving force (incident radiation), advection of energy
to the nightside (high velocity winds), nightside heating (through
compression, heating, shocks, and other sources of viscosity), and
finally re-radiation on the nightside \citep{goodman2009}. To first
order a steady-state solution is found in the models of most other
groups (see Section 1 for references) including those that use other
gridding techniques that do not necessitate any polar
approximations. The second reason our approximation appears valid is
any temporal variation that is seen in simulations is primarily
confined to latitudes that are equator-ward of $\approx
60^{\circ}$. These motions are unlikely to be influenced by slightly
shorter timesteps taken at the poles. Finally, the third justification
of our polar treatment is a series of runs with varying $\theta_{max}$
from $70^{\circ}$ to $90^{\circ}$. Comparing the zonal (longitudinal)
velocity calculated these runs we find that increasing $\theta_{max}$
widens the latitudinal extent of the zonal jet and causes a slight
shift in peak velocity. However, the results appear to have largely
converged by $\theta_{max}=85^{\circ}$. For the remainder of the paper
we set $\theta_{max}=85.5^{\circ}$.

\subsubsection{Wavelength-Dependent Stellar Energy Deposition}
We have significantly improved the method by which we calculate where
the stellar energy is deposited into the upper atmosphere by utilizing
a wavelength-dependent prescription rather then a gray
approach. Following \citet{showman2009}, we divide the spectra into
$30$ wavelength bins (see \citet{showman2009} Table 1 for a list of
the bin boundaries) and calculate the opacity within each bin
independently. We then calculate the spatial distribution of energy
deposition in each of these bins and sum over all bins to get the net
energy deposition at each location. The equation for the thermal
energy can be written as
\begin{eqnarray}
\left[ \pdt{\epsilon} + ({\bf u}\cdot\nabla) \epsilon \right] = - P
\nabla \cdot {\bf u} - \rho \kapp\left[\bbody - c\er \right]
\\ \nonumber + D_v + \left(\frac{R_{\star}}{a}\right)^2 \rho \sum_{b=1}^{30}\pi
B\left(T_{\star}\right)_b  \kappa_b e^{-\tau_b/\mu} \Delta\nu.
\label{eq:thermalenergy}
\end{eqnarray}
The second term on the left represents the advection of thermal energy
and the terms on the right are the work done by compression, the
exchange of energy between matter and radiation, viscous heating, and
the direct heating by incident stellar irradiation, respectively. Note
that we have also added a $\mu=\cos\left(\theta_{\star}\right)$, where
$\theta_{\star}$ is the angle between the local normal and the
incident radiation. This accounts for the additional material stellar
photons encounter when traversing the limb of the planet (For a more
detailed description of the derivation of this equation and the
radiation energy equation see \citet{dobbsdixon2010}).

\begin{figure}
\plotone{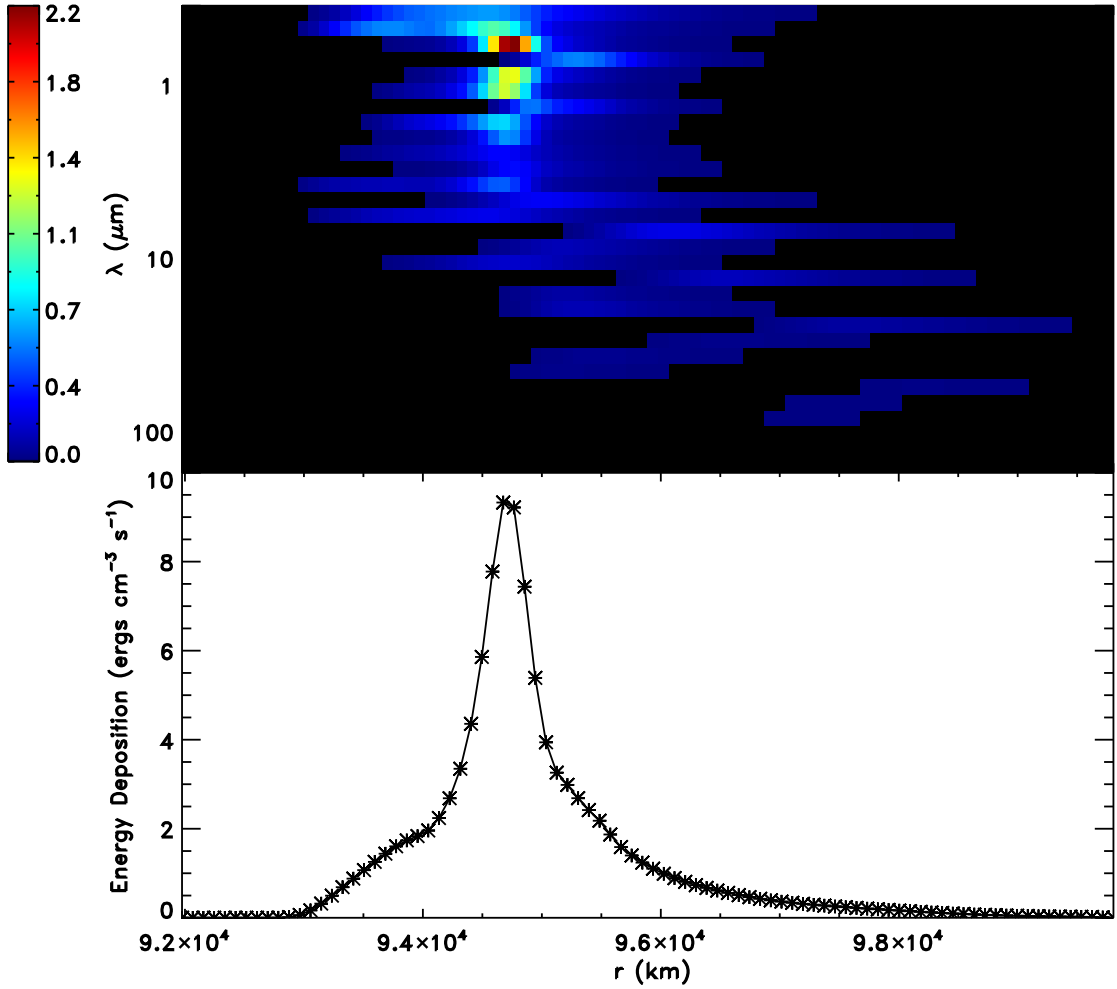}
\caption{The behavior of the stellar energy deposition term
  $\left(\frac{R_{\star}}{a}\right)^2 \pi B_b\left(T_{\star}\right)
  \rho\kappa_b e^{-\tau_b/\mu} \Delta\nu$ from Equation
  (\ref{eq:thermalenergy}) at the substellar point for each wavelength
  bin (upper panel) and as summed over all bins (lower panel).}
\label{fig:stellarheating}
\end{figure}

In Figure (\ref{fig:stellarheating}) we show the behavior of the last
term in Equation (\ref{eq:thermalenergy}) at the sub-stellar point. The
upper panel illustrates heating as a function of both depth and
wavelength, while the lower panel shows the sum over all the
wavelength bins. We find that the distribution and overall magnitude of
energy deposition using this binning method matches a full
wavelength-dependent calculation quite well. When compared to the gray
approach of \citet{dobbsdixon2010}, we find the region of energy
deposition is somewhat more extended. Regions of the spectra with low
opacity (near $3 \mu m$ for example) are able to penetrate further
into the atmosphere.

\subsubsection{Flux-Limited Diffusion Boundary Conditions}
The classic form of flux-limited diffusion (FLD) \citep{levermore1981}
postulates that the radiative flux can be written as ${\bf F} = -
\lambda c/\left(\rho\kapr\right) \nabla \er$, where $\lambda$ is a
temporally and spatially variable flux limiter providing the closure
relationship between flux and radiation energy density. It is
typically formulated to give limits for flux in the optically-thick
regime, ${\bf F} = -4acT^3/\left(3\rho\kapr\right) \nabla T$, and the
optically-thin regime, $|{\bf F}| = c\er$.

The problem lies in the application of the free-streaming,
optically-thin limit to planetary atmospheres. The free-streaming
limit assumes that all of the radiation is collimated and moving in
the outward direction at $\tau=0$.  Thus, by conservation of energy,
$\er=Fc$, {\it i.e.} the energy density in any given cell is
determined by the rate at which radiation flows through that cell,
assuming Cartesian geometry.

However, this is {\bf not} the limit we are in at the photosphere of
the planet.  Since limb-darkening of the planet is weak, an observer
placed close to $\tau=0$ near the surface of the planet will see
thermal radiation that is nearly isotropic from $2\pi$ steradians
towards the planet and no thermal radiation from the opposite
hemisphere. Therefore, the free-streaming limit does not apply close
to the planet. Instead, we must consider the energy density and flux
given the non-isotropic intensity. The energy density is the first
moment of the intensity and can be written $\er = \int_{2\pi}I_0
d\Omega = 2\pi I_0$. The flux is simply the second moment and can be
expressed as $F=c\int_{2\pi}I_0\mu d\Omega = c\pi I_0$. Comparing
these expressions we see that the proper relation between energy
density and flux at the boundary should be $F\sim c\er/2$. A given
flux, set by the sum of the incident and internal flux, will result in
a larger energy density, and equivalently a higher temperature.

\subsection{Radiative Transfer}
To determine the transit spectra of HD209458b models, we calculate the
wavelength dependent absorption of stellar light traversing through
the limb of the planet. This allows us to determine the effective
radius of the planet and a fractional reduction of stellar flux
$F_{\star}$. Neglecting limb-darkening of the star, this can be
expressed as
\begin{equation}
\left(\frac{F_{in-transit}}{F_{\star}}\right)_{\lambda} = \frac{\int
  \left(1-e^{-\tau\left(b,\alpha,\lambda\right)}\right)b db d\phi}{\pi
  R_{\star}^{2}},
\label{eq:fluxratio}
\end{equation}
where $\tau\left(b,\alpha,\lambda\right)$ is the total optical depth
along a given chord with impact parameter b and polar angle $\alpha$,
defined on the observed planetary disk during transit. The density and
temperature needed to calculate $\tau$ at each location are
interpolated from the values in the 3D models.

Wavelength-dependent opacities are taken from the atomic and molecular
opacity calculations of \citet{sharp2007}. These calculations neglect
the effects of grains on absorption, assuming that grains will rain
out of the upper atmosphere before they grow to any significant
size. 

\section{Calculated Transit Spectra}
In this section we present transit spectra calculated from 3D
radiative-hydrodynamical models exploring the different flow
structures among models with varying viscosity. While several features
may influence the transit spectra at precisions already observed,
others must await the next generation of instruments (as discussed in
the conclusion). We highlight three methods for extracting this signal
from actual data, exploiting the differences between transit spectra
taken during transit ingress and egrees. We show that the time
dependence of the transit signal can be used to extract information
about the eastern and western hemispheres.

\subsection{Varying Viscosity}
The source of viscosity in the atmospheres of irradiated giant planets
is a major outstanding issue \citep{li2010_2} and can play an
important role in shaping the overall structure of the atmospheric
dynamics and energy re-distribution. Physically motivated sources of
viscosity can arise from both unresolved process (sub-grid effects) or
missing physics ({\it e.g.} magnetic viscosity as discussed by
\citet{batygin2010} and \citet{perna2010}). Unresolved processes may
include the generation of turbulence through shocks, instabilities, or
waves. Numerical viscosity, present in all numerical simulations, may
also allow flow to smoothly traverse additional shocks
throughout the simulation. Such shocks may, if resolved, act as an
additional source of viscosity converting the kinetic energy of the
jet to thermal energy that can be subsequently radiated away.

\begin{figure}
\plotone{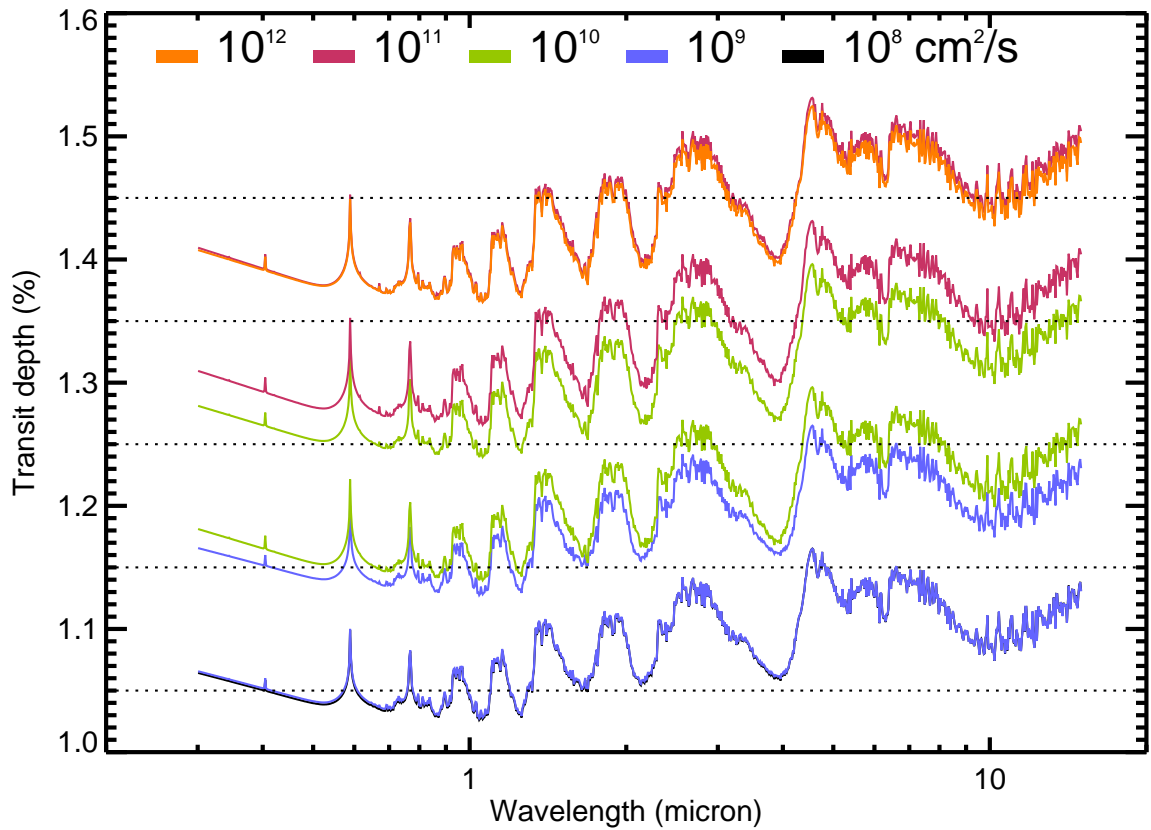}
\caption{Difference in mid-transit transit spectra among models
  with varying viscosity. Successive pairs of spectra, with decreasing
  viscosities, are shifted downward $0.1\%$ for clarity. The reference
  dotted lines are similarly offset.}
\label{fig:viscous}
\end{figure}

To explore the role of an isotropic viscosity disregarding its
potential origin, \citet{dobbsdixon2010} performed a series of
simulations with varying kinematic viscosity ($\nu$). We have
revisited these models allowing for the cross-pole flow as described
in Section (2.1.1), wavelength-dependent heating (2.1.2), an improved
FLD boundary condition (2.1.3), and significantly higher spatial
resolution. Transit spectra calculated from our models during
mid-transit are shown in Figure (\ref{fig:viscous}). The variations
due to varying flow structures (See Figures (2) and (3) of
\citet{dobbsdixon2010} for an illustration of the differences) can be
quite dramatic, including a transition from subsonic to supersonic
wind speeds as the viscosity is lowered. Table (\ref{table:one}) gives
a summary of the simulation parameters.

\begin{table}
\begin{center}
\begin{tabular}{|c|c|c|c|c|}
\hline
Simulation & $\nu$ ($\mathrm{cm^2/s}$) & $\alpha_{eff,ph}$ & $H_{ss,P}$ (km) & peak $v_{\phi}$ (km/s)\\
\hline
\hline
$S1$ & $10^{12}$ & $10^{-1}$ & 538 & 0.76 \\
\hline
$S2$ & $10^{11}$ & $10^{-2}$ & 501 & 1.48 \\
\hline
$S3$ & $10^{10}$ & $10^{-3}$ & 489 & 3.70 \\
\hline
$S4$ & $10^{9}$ & $10^{-4}$ & 481 & 4.90 \\
\hline
$S5$ & $10^{8}$  & $10^{-5}$ & 461 & 4.82 \\
\hline
\end{tabular}
\end{center}
\caption{Values of kinematic viscosity used for the simulations
  presented here. For reference we also quote an average effective
  alpha-parameter, sub-stellar pressure scale-height at the
  photosphere, and peak zonal velocities, increasing with decreasing
  viscosity.}
\label{table:one}
\end{table}

The transit spectra shown in Figure (\ref{fig:viscous})
illustrate a number of interesting features. Common to all spectra are
broad absorption features due primarily to H$_2$O, CH$_4$, and CO,
strong narrow features due to Na and K, and a rising signal at
short wavelengths associated with Rayleigh scattering. The highest
viscosity simulation has the largest overall transit depth at all
wavelengths. The overall transit depth decreases roughly monotonically
with decreasing viscosity, though spectra for the two highest and two
lowest viscosities are virtually indistinguishable.

The trend of increasing transit depth with viscosity can be understood
by examining the transition in the dynamics as viscosity is
decreased. The highest viscosity simulations significantly restricts
advection across the terminators. This results in symmetric flow
across the eastern and western terminators, both with average
scale-heights of approximately $450$ km. The zonal flow is subsonic
and dies out soon after crossing the terminator. However, as the
viscosity is decreased a super-rotating jet develops that
circumnavigates the entire planet. This fundamentally changes the
temperature of the advected gas across the terminators. For the
high-viscosity simulations, the flow across both terminators is
advecting gas from day to night. For the lowest viscosity simulation,
the circumplanetary jet implies that the flow is still advecting gas
from day to night at the eastern terminator (defined relative to the
sub-stellar point), but from night to day at the western
terminator. This results in scale-heights of $440$ km at the eastern
terminator, but only $340$ km at the western terminator (we exploit
this difference in the next sub-sections). Mid-transit spectra
represent the average of the two, leading to larger overall transit
depths for high-viscosity simulations.

The transition from symmetric to antisymmetric flow patterns can also
be understood within the context of the work of
\citet{showman2011}. The transport of angular momentum up-gradient to
form an equatorial super-rotating circumplanetary jet is achieved
primarily horizontally through eddies that have northwest/southeast
(southwest/northeast) phase tilt in the northern (southern)
hemisphere. The result is that eddies in both hemispheres transfer
eastward angular momentum toward the equator and westward angular
momentum to higher latitudes. We have performed eddy-analysis on our
results utilizing the formalism of \citet{karoly1998} and confirm that
meridional eddies are the dominate source of the equatorial jet's
angular momentum. The primary excitation of these eddies are zonally
propagating planetary-scale (Rossby and Kelvin) waves. High-viscosity
simulations effectively inhibit the propagation of these waves and
their associated eddies and thus preclude the formation of
super-rotating circumplanetary jets.

\subsection{Extracting the Transit shape}
\begin{figure}
\plotone{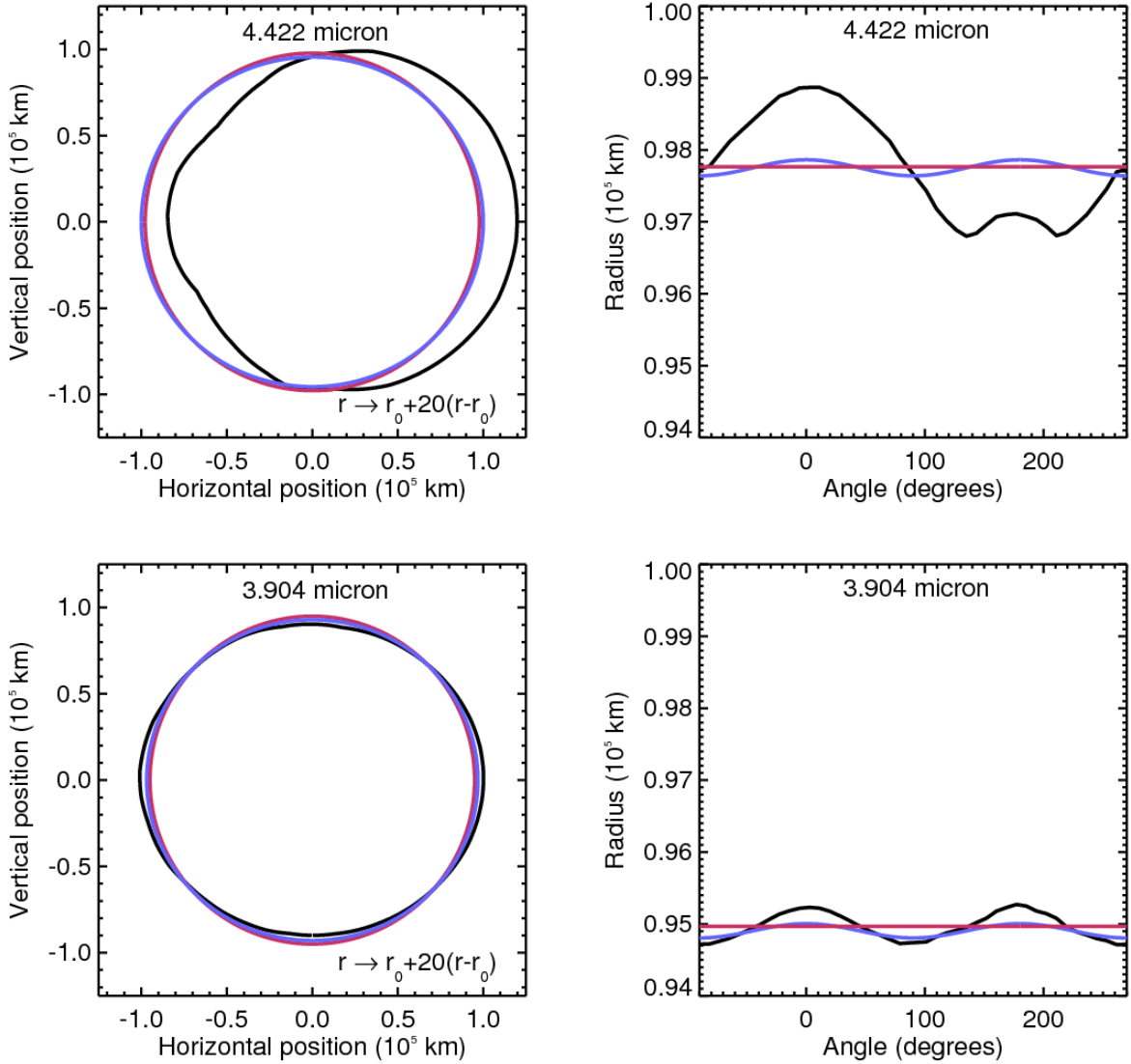}
\caption{The wavelength dependent shape of the planet at $4.422\mu m$
  (top row) and $3.904\mu m$ (bottom row) for the $\nu=10^{8}$ cm$^2$
  s$^{-1}$ simulation. The left column shows the shape of the planet
  as seen during transit. The black curves illustrates the actual
  planetary shape (where $\tau\left(\lambda\right)=2/3$), the red
  curve illustrates a spherical planet, and the blue curve illustrates
  a planet filling a Roche-potential (both normalized to have the same
  total area). The planet shape has been exaggerated by a factor of 20
  in the left column. The right column shows the same quantities as a
  function of latitude along the day-night delineator. An angle of $0$
  corresponds to the equator and $\pi/2$ to the northpole.}
\label{fig:shape}
\end{figure}

For number of reasons, including rotation
\citep{seager2002,barnes2003}, strong zonal winds \citep{barnes2009},
equator-pole temperature gradients, and the tidal potential of the
host star \citep{leconte2011} short period, synchronously rotating,
irradiated planets are not spherical. In order to explore these
variations we illustrate the shape of the planet at $4.422$ and $3.904
\mu m$ in the black curve of Figure (\ref{fig:shape}) for the
simulation with the lowest viscosity ($10^8 \mathrm{cm^2/s}$). As can
be seen in Figure (\ref{fig:viscous}), these two wavelengths were
chosen to have the largest ($4.422 \mu m$) and smallest ( $3.904 \mu
m$) transit light depths for wavelengths shortward of $10\mu m$. This
implies that $3.904 \mu m$ observations will probe much deeper into
the atmosphere where the planet becomes more spherically
symmetric. Higher in the atmosphere at $4.422 \mu m$, there are
significant discrepancies between the mean surface (shown in red) and
the photosphere, illustrating the dominant role of energy
redistribution in determining the detailed shape of the planet. Below,
we use this fact to justify extracting detailed temperature
distributions from variations in the planetary shape with wavelength.

Also shown in Figure (\ref{fig:shape}) is the geopotential surface (in
blue) corresponding to the mean radius (in red). The geopotential
combines the gravity and centrifugal forces into a single potential
$\nabla Q = 1/2\nabla\left({\bf\Omega}\times{\bf r}\right)^2 +
\nabla\Phi$. For a slow rotating planet such as HD209458b the
temperature gradients across the planet (yielding varying
scale-heights) are the largest factor in changing the planets
shape. This can be seen most clearly at $3.904 \mu m$ where, despite
having the least amount of asymmetry, the oblateness caused by
scale-height changes (black line) exceeds the rotational oblateness of
the planet (red line).

Because the planet's absorption cross-section is oblate, due primarily
to the high temperatures, we can utilize observational information of
the transit shape to study the efficiency of dynamical redistribution
of energy throughout the atmosphere. In addition to the equator-pole
differences, in models with low viscosity that have strong
super-rotating recirculation, the eastern terminator has at a higher
temperature than the western, and thus has a larger scale-height, so
the eastern hemisphere presents a larger absorption cross-section than
the western hemisphere. This is also evident in the black curve of
Figure (\ref{fig:shape}). Consequently, the planet has an asymmetric
egg-shaped absorption cross-section which directly translates into an
asymmetry in the transit shape. In the remainder of this subsection we
present three methods for diagnosing this observationally:
wavelength-dependent transit timing, differences between ingress and
egress spectra, and color-dependent transit shape.

\begin{figure}
\plotone{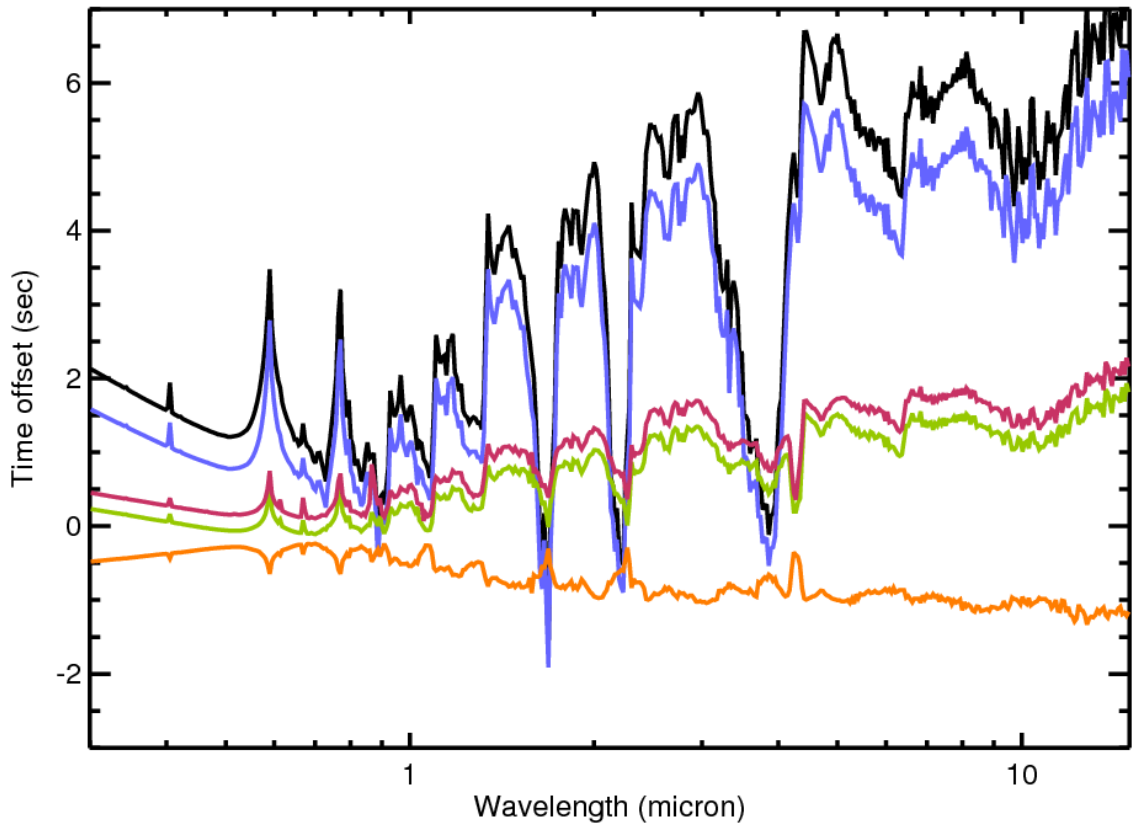}
\caption{Effective time offset versus wavelength derived by forcing a
  circular fit to the planetary shape. Timing variations are primarily
  due to differences in the temperature structure between eastern and
  western terminators. Colors represent simulations with varying
  viscosities as in Figure (\ref{fig:viscous}).}
\label{fig:timeoffset}
\end{figure}

The first method we discuss is wavelength-dependent transit
timing. Since the western terminator has a smaller scale-height and
this part of the planet transits first, the ingress will be slightly
delayed, while the eastern terminator is more extended, causing the
end of ingress to be delayed. Likewise, the egress will be delayed as
well, so the overall shift will be delay the transit relative to the
center of mass of the planet. At wavelengths with larger opacity this
asymmetry is stronger so the transit time delay is larger, while at
wavelengths with smaller opacity it is weaker; consequently, the
central time of transit will appear to vary with wavelength if one
fits the transit with a symmetric planet model. Figure
\ref{fig:timeoffset} shows the effective transit time offset versus
wavelength computed for a model with no limb-darkening for the star
and by fitting the transit at each wavelength with a circular planet
model \citep{mandel2002}. As the viscosity grows, the two hemispheres
have smaller temperature differences, and hence, the transit shape is
more symmetric, causing a smaller time offset.

\begin{figure}
\plotone{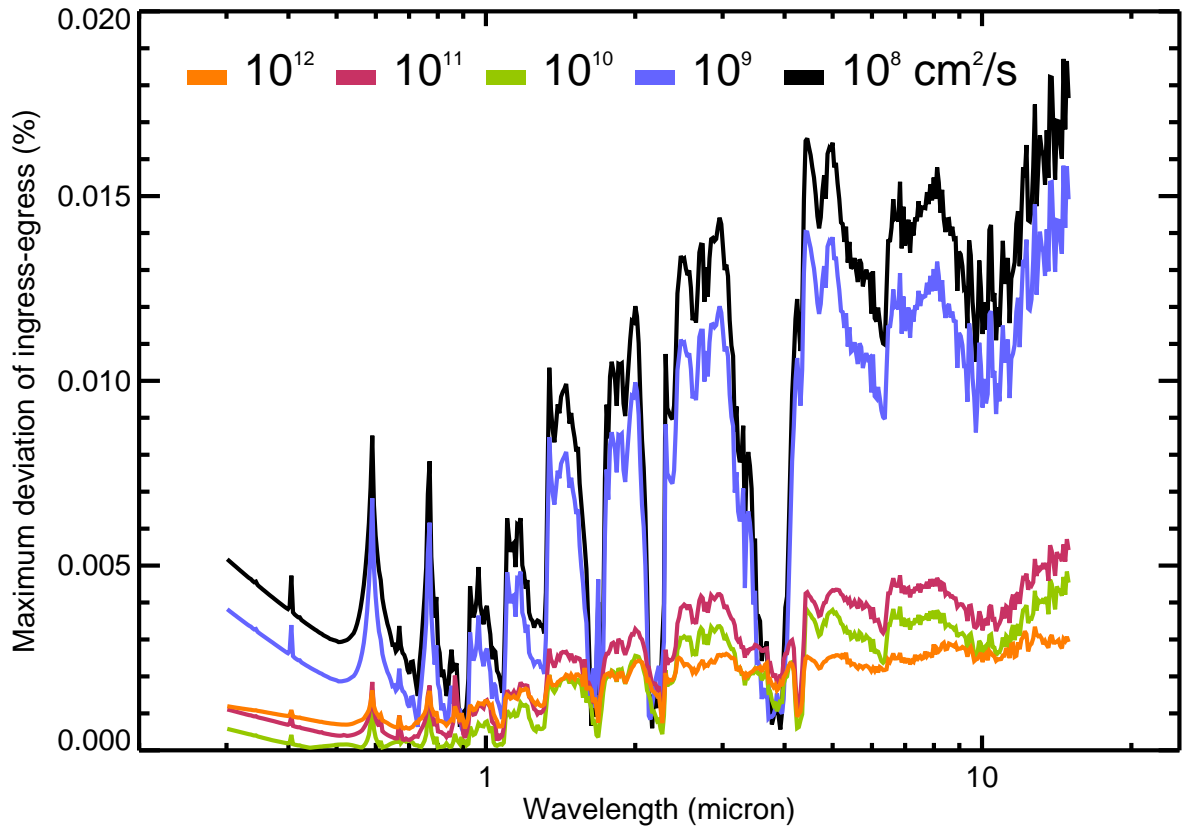}
\caption{Maximum deviation between the ingress and egress spectra. To
  calculate these curves we subtract the absolute value of the
  lightcurve from the same lightcurve time-reversed about the central
  time of transit.}
\label{fig:maxdev}
\end{figure}

A second means of diagnosing the difference between the hemispheres is
to flip the light curve about the midpoint and subtract it from the
original. The deviations that remain show the difference between the
ingress and egress caused by a difference in shape of the east and
west terminators, in addition to the time offset. To characterize
this, we compute the maximum value of this deviation for each light
curve computed at a specific wavelength. Figure \ref{fig:maxdev} shows
this maximum deviation versus wavelength. To compute this, we
calculate the maximum of the absolute value of the lightcurve minus
the time-reversed light curve. The largest deviations are at the
$10^{-4}$ level for wavelengths dominated by water absorption opacity,
and these deviations grow as the viscosity decreases due to the larger
differences between the terminators for smaller viscosity. In
practice, finding the midpoint of the transit around which to flip the
light curve is not straightforward. As noted by \citet{barnes2009}, as
transit photometry is sensitive to the shape of the planet and not the
center of mass, the offset between the two cannot be determined {\it
  ab initio}. However, observing at multiple wavelengths, one may
utilize the wavelength with the smallest transit depth to define the
center of transit with the understanding that this is probing the
deepest, most spherically symmetric region in the planet. This is
illustrated in Figure (\ref{fig:shape}).

\begin{figure}
\plotone{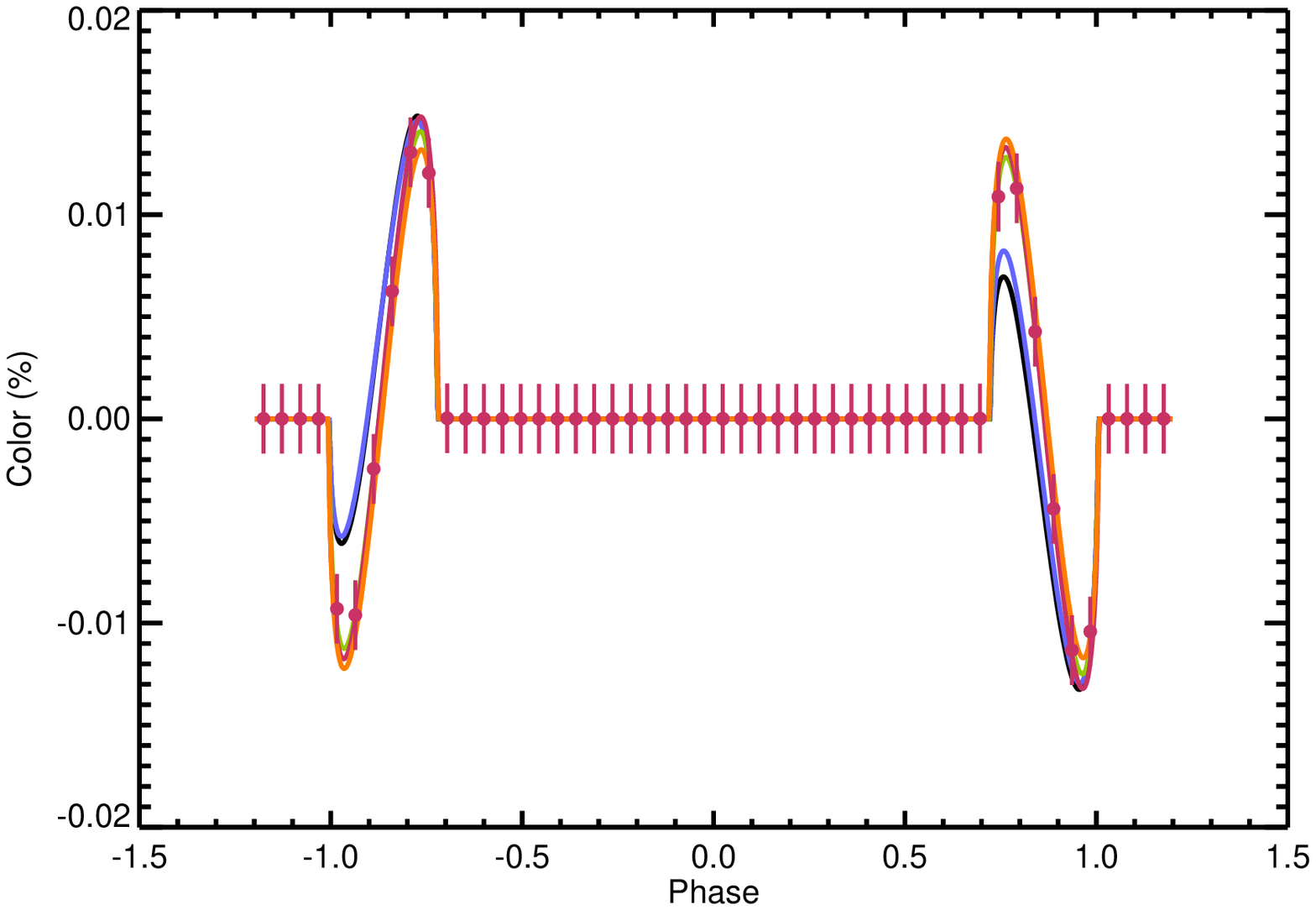}
\caption{Color of transit for the F162M minus the re-normalized F277W
  JWST NIRCam filters. Vertical bars indicate predicted errors for
  NIRCam.}
\label{fig:color}
\end{figure}

Finally, a third, model-independent diagnostic for the planet
asymmetry is the color-dependent transit shape. Figure \ref{fig:color}
shows the difference in the shape of transit for two wave bands:
1.55-1.70 and 2.4-3.1 micron, corresponding to the proposed JWST
NIRCam F162M and F277W filters, respectively. To compute the shape
difference, we subtract the two lightcurves after renormalizing the
depth of the second to match the depth of the first. This
model-independent measurement of the difference in the shape of the
transit in these two wave-bands can be written as
$C=(D_1-D_2D_{1,max}/D_{2,max})$ where $D_{1,2}$ are the depths of
transit (in dimensionless units the out-of-transit flux minus the
in-transit flux divided by the out-of-transit flux) and the {\it max}
subscript indicates the maximum depth of transit. The asymmetry in
these lightcurves clearly demonstrates that at different wavelengths
the planet absorbs with a different asymmetric cross section. These
filters have several advantages; they can be observed simultaneously
with JWST using the dichroic and the two bands are centered at
wavelengths with high and low water opacity, respectively, yielding a
larger differential signal.

\subsection{Comparison to observations}
As a check on the total depth of transit, we have compared the transit
depth to observations by \citet{knutson2008} and
\citet{beaulieu2010}. These observations cover wavelengths from $0.3$ to
$10$ microns, and thus provide an important validation test of these
numerical models. We have averaged the models over the observed bands,
and have varied the value of the inner radius of the simulation zone
to obtain the best agreement with the observed transit depths.  This
is not as accurate as carrying out additional simulations in which the
model region is varied; however, this would be much more
computationally expensive. Furthermore, the assumption of constant
{\bf g} in both these simulations and those of all other groups allows
for such a shift as only the curvature terms are affected. The applied
depth offset is quite small, amounting to a maximum correction of only
a $1\%$ to the planetary radius. Finally, we exclude the bands with
strong sodium and potassium absorption as there is evidence that
potassium is depleted (which is not accounted for in our opacity
tables).

\begin{figure}
\plotone{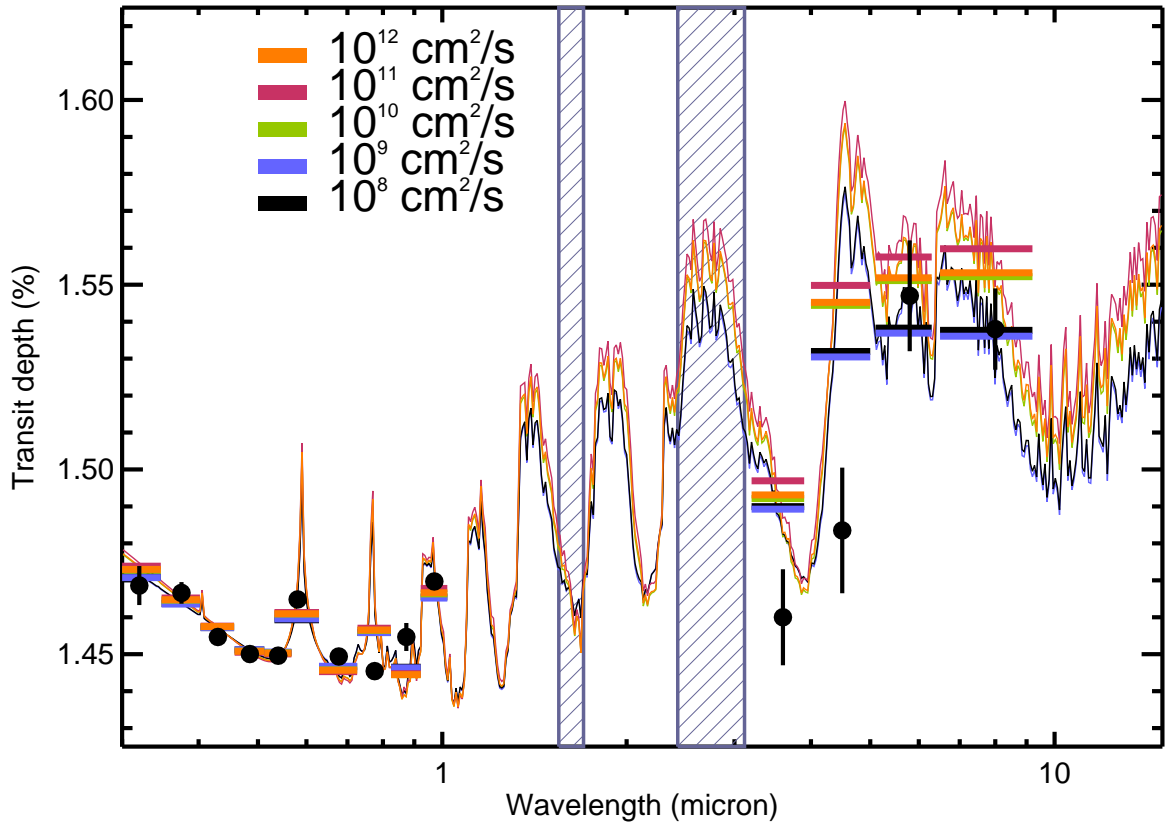}
\caption{Transit depth versus wavelength. The black dots with error
  bars are the data from \citet{knutson2007} and \citet{beaulieu2010}.
  The thick horizontal bars are the four models averaged over the
  measured wave bands. The curves show model spectra binned by a
  factor of 10 in wavelength.}
\label{fig:transitdepth}
\end{figure}

There are twelve wave bands that we use, and with one free parameter,
gives eleven degrees of freedom for each model. Figure
\ref{fig:transitdepth} shows the comparison of the models to the
data. We find best-fit $\chi^2$ of $25.41$, $25.04$, $37.62$, $45.73$,
and $38.77$ for the models with viscosities of $10^8, 10^9, 10^{10}, 10^{11}$
and $10^{12}$ cm$^2$ s$^{-1}$ respectively. Qualitatively, the
overall agreement of the models with the data is quite good: (1) the
transit depth is weakly dependent on the viscosity; (2) there are no
discrepancies between the data and model greater than $2-\sigma$; (3)
the observed transit spectrum with wavelength shows the expected
features due to water and Rayleigh scattering. In general, the success
of this model is comparable to transit spectra calculated from
other models which utilize GCM simulations
\citep{fortney2010,burrows2010}.

However, in detail there are significant discrepancies; in particular,
the observed IRAC transit depths appear to vary more strongly with
wavelength than the model predicts. This is reflected in the larger
$\chi^2$ of the fits, of which one-half to two-thirds is due to the
infrared discrepancies. The fit to the infrared data obtained by
\citet{beaulieu2010} has an extremely good $\chi^2$; however, their
model was one-dimensional, and allowed the abundances and
temperature/pressure profile to float, so it is not surprising that
they obtain a good fit with so many degrees of freedom. Another
possibility is that systematic errors still exist in the IRAC data
reduction. For example, for the transiting planet HD 189733b, different
groups have obtained markedly different transit depths at infrared
wavelengths using the same IRAC data sets; consequently, there may be
some remaining systematic error present in the data
\citep{desert2009}. The final possibility is that there is still
physics that is not included in our models which is causing the
discrepancies; for example, varied chemical abundances,
non-equilibrium chemistry, and magnetic drag have not been included in
these models.

\section{Discussion}
In this paper we have presented wavelength-dependent transit spectra
for a series of simulations utilizing a state of the art, 3D radiative
hydrodynamical model. The simulations presented here have been updated
from \citet{dobbsdixon2010} to allow for flow over the polar regions,
wavelength-dependent stellar energy deposition which accounts for the
slant optical depth, and an improved boundary condition for
flux-limited diffusion. By `observing' these simulations we are able
to explore potentially detectable diagnostics of the overall state of
the atmospheric dynamics.

We have explored the role of changing viscosity on the observable
transit spectra. Larger viscosity (potentially caused by
unresolved instabilities, shocks, magnetic fields, etc.) significantly
alters the overall flow structure. These changes lead to variations in
the observed transit spectra during the center of primary
transit, but are much more evident in differential spectral
measurements. We explore three separate methods to detect such
variation: wavelength-dependent transit timing, differences in
absorption between ingress and egress, and color-dependent transit
shape. We also explored the the observable variations in transit
spectra due to exo-weather. \citet{dobbsdixon2010} identified
dynamically-induced variation in their intermediate viscosity
simulation. Based on these results and the range of observational
evidence both for and against variability
\citep{grillmair2008,agol2010} we found sufficient motivation to look
for signatures of variability. However, we find that the observable
variations in transit spectra due to exo-weather are somewhat
limited and likely undetectable, even with JWST. However, one final
note of caution is in order. The weather related phenomena we explore
are limited to dynamically induced variations. Chemically-induced
variations due to clouds, non-equilibrium chemistry, etc. may
potentially couple positively to dynamical variations and amplify the
observability of exo-weather.

We have compared our analytical results for simulations with varying
viscosity to the observations of HD209458b by \citet{knutson2007} and
\citet{beaulieu2010}. In general, we find quite good agreement, with
none of the discrepancies greater then $2-\sigma$. Overall, our best
fit model is the simulation with viscosity set to $10^9
\mathrm{cm^2/s}$, which has developed a super-rotating circumplanetary
jet at the equator. The largest differences between our models and the
observations occur in the IR where we find that we consistently
over-predict the transit depth in the observed $3.6$ and $4.5\mu m$
IRAC bands. \citet{fortney2010} and \citet{burrows2010} have performed
an analysis similar to ours utilizing GCM dynamical models. They find
disagreements with the IRAC bands as well, over-predicting the shorter
bands and under-predicting the longer IRAC bands. \citet{shabram2011}
also discusses difficulties matching the models of
\citet{beaulieu2010} to analytic models of \citet{desetangs2008}.

Although the measurements presented here are likely beyond our current
capabilities it is informative to speculate on future
possibilities. The quoted error bars in \citet{beaulieu2010} are
$0.013$, $0.017$, $0.015$ and $0.011$ for the four IRAC wavebands
(with increasing $\lambda$). The James Webb Space Telescope (JWST) will
have approximately $43$ times the collecting area of the Spitzer
telescope. Assuming that the S/N is photon-limited, this will yield
errors for the full transit that are $6.6$ times smaller. A similar
measurement would thus have errors of approximately $0.0022\%$. Given
the results presented here, distinguishing between models with varying
viscosity is potentially accessible to JWST observations.

Utilizing the difference between ingress and egress spectra through
one of the methods described here remains the most viable mechanism to
distinguish between dynamical models. Errors in this measurement
increase due to the decreased time of observation. Ingress or egress
would be shorter than the transit duration by $\frac{R_p}{2
  R_{\star}}\frac{1}{1-b^2} \approx 0.08$, where $b$ is the impact
parameter of the orbit. This gives total errors in a JWST measurement
of the order of $0.0076$. Figure (\ref{fig:maxdev}) shows
wavelength-dependent deviations are detectable given this level of
error. The color dependent transit light curve shown in Figure
(\ref{fig:color}) explicitly shows the expected error, again
illustrating the clear detectability of this signal.

\acknowledgements IDD is supported by the Carl Sagan Postdoctoral
program, EA was supported in part by the Miller Institute for Basic
Research in Science and by the National Science Foundation under
CAREER Grant No.\ 0645416, and AB acknowledges support in part under
NASA ATP grant NNX07AG80G, HST grants HST-GO-12181.04-A and
HST-GO-12314.03-A, and JPL/Spitzer Agreements 1417122, 1348668,
1371432, and 1377197. This work was partly completed at the Kavli
Institute for Theoretical Physics at Santa Barbara. We would also like
to acknowledge the use of NASA's High End Computing Program computer
systems.

\bibliographystyle{aa}
\bibliography{ian}

\begin{thebibliography}{53}
\expandafter\ifx\csname natexlab\endcsname\relax\def\natexlab#1{#1}\fi

\bibitem[{{Agol} {et~al.}(2010){Agol}, {Cowan}, {Knutson}, {Deming}, {Steffen},
  {Henry}, \& {Charbonneau}}]{agol2010}
{Agol}, E., {Cowan}, N.~B., {Knutson}, H.~A., {et~al.} 2010, \apj, 721, 1861

\bibitem[{{Barnes} {et~al.}(2009){Barnes}, {Cooper}, {Showman}, \&
  {Hubbard}}]{barnes2009}
{Barnes}, J.~W., {Cooper}, C.~S., {Showman}, A.~P., \& {Hubbard}, W.~B. 2009,
  \apj, 706, 877

\bibitem[{{Barnes} \& {Fortney}(2003)}]{barnes2003}
{Barnes}, J.~W. \& {Fortney}, J.~J. 2003, \apj, 588, 545

\bibitem[{{Batygin} \& {Stevenson}(2010)}]{batygin2010}
{Batygin}, K. \& {Stevenson}, D.~J. 2010, ArXiv e-prints

\bibitem[{{Beaulieu} {et~al.}(2010){Beaulieu}, {Kipping}, {Batista}, {Tinetti},
  {Ribas}, {Carey}, {Noriega-Crespo}, {Griffith}, {Campanella}, {Dong},
  {Tennyson}, {Barber}, {Deroo}, {Fossey}, {Liang}, {Swain}, {Yung}, \&
  {Allard}}]{beaulieu2010}
{Beaulieu}, J.~P., {Kipping}, D.~M., {Batista}, V., {et~al.} 2010, \mnras, 409,
  963

\bibitem[{{Brown} {et~al.}(2002){Brown}, {Libbrecht}, \&
  {Charbonneau}}]{brown2002}
{Brown}, T.~M., {Libbrecht}, K.~G., \& {Charbonneau}, D. 2002, \pasp, 114, 826

\bibitem[{{Burkert} {et~al.}(2005){Burkert}, {Lin}, {Bodenheimer}, {Jones}, \&
  {Yorke}}]{burkert2005}
{Burkert}, A., {Lin}, D.~N.~C., {Bodenheimer}, P.~H., {Jones}, C.~A., \&
  {Yorke}, H.~W. 2005, \apj, 618, 512

\bibitem[{{Burrows} {et~al.}(2010){Burrows}, {Rauscher}, {Spiegel}, \&
  {Menou}}]{burrows2010}
{Burrows}, A., {Rauscher}, E., {Spiegel}, D.~S., \& {Menou}, K. 2010, \apj,
  719, 341

\bibitem[{{Burrows} {et~al.}(2003){Burrows}, {Sudarsky}, \&
  {Hubbard}}]{burrows2003}
{Burrows}, A., {Sudarsky}, D., \& {Hubbard}, W.~B. 2003, \apj, 594, 545

\bibitem[{{Charbonneau} {et~al.}(2000){Charbonneau}, {Brown}, {Latham}, \&
  {Mayor}}]{charbonneau2000}
{Charbonneau}, D., {Brown}, T.~M., {Latham}, D.~W., \& {Mayor}, M. 2000, \apjl,
  529, L45

\bibitem[{{Charbonneau} {et~al.}(2002){Charbonneau}, {Brown}, {Noyes}, \&
  {Gilliland}}]{charbonneau2002}
{Charbonneau}, D., {Brown}, T.~M., {Noyes}, R.~W., \& {Gilliland}, R.~L. 2002,
  \apj, 568, 377

\bibitem[{{Cho} {et~al.}(2003){Cho}, {Menou}, {Hansen}, \& {Seager}}]{cho2003}
{Cho}, J.~Y.-K., {Menou}, K., {Hansen}, B.~M.~S., \& {Seager}, S. 2003, \apjl,
  587, L117

\bibitem[{{Cho} {et~al.}(2008){Cho}, {Menou}, {Hansen}, \& {Seager}}]{cho2008}
---. 2008, \apj, 675, 817

\bibitem[{{Cooper} \& {Showman}(2005)}]{cooper2005}
{Cooper}, C.~S. \& {Showman}, A.~P. 2005, \apjl, 629, L45

\bibitem[{{Cooper} \& {Showman}(2006)}]{cooper2006}
---. 2006, \apj, 649, 1048

\bibitem[{{Courant} {et~al.}(1928){Courant}, {Friedrichs}, \&
  {Lewy}}]{courant1928}
{Courant}, R., {Friedrichs}, K., \& {Lewy}, H. 1928, Mathematische Annalen,
  100, 32

\bibitem[{{Cowan} \& {Agol}(2008)}]{cowan2008}
{Cowan}, N.~B. \& {Agol}, E. 2008, ArXiv e-prints

\bibitem[{{Cowan} {et~al.}(2007){Cowan}, {Agol}, \& {Charbonneau}}]{cowan2007}
{Cowan}, N.~B., {Agol}, E., \& {Charbonneau}, D. 2007, \mnras, 379, 641

\bibitem[{{Crossfield} {et~al.}(2010){Crossfield}, {Hansen}, {Harrington},
  {Cho}, {Deming}, {Menou}, \& {Seager}}]{crossfield2010}
{Crossfield}, I.~J.~M., {Hansen}, B.~M.~S., {Harrington}, J., {et~al.} 2010,
  \apj, 723, 1436

\bibitem[{{Deming} {et~al.}(2005){Deming}, {Seager}, {Richardson}, \&
  {Harrington}}]{deming2005}
{Deming}, D., {Seager}, S., {Richardson}, L.~J., \& {Harrington}, J. 2005,
  \nat, 434, 740

\bibitem[{{D{\'e}sert} {et~al.}(2009){D{\'e}sert}, {Lecavelier des Etangs},
  {H{\'e}brard}, {Sing}, {Ehrenreich}, {Ferlet}, \&
  {Vidal-Madjar}}]{desert2009}
{D{\'e}sert}, J.-M., {Lecavelier des Etangs}, A., {H{\'e}brard}, G., {et~al.}
  2009, \apj, 699, 478

\bibitem[{{Dobbs-Dixon} {et~al.}(2010){Dobbs-Dixon}, {Cumming}, \&
  {Lin}}]{dobbsdixon2010}
{Dobbs-Dixon}, I., {Cumming}, A., \& {Lin}, D.~N.~C. 2010, \apj, 710, 1395

\bibitem[{{Dobbs-Dixon} \& {Lin}(2008)}]{dobbsdixon2008_1}
{Dobbs-Dixon}, I. \& {Lin}, D.~N.~C. 2008, \apj, 673, 513

\bibitem[{{Fortney} {et~al.}(2010){Fortney}, {Shabram}, {Showman}, {Lian},
  {Freedman}, {Marley}, \& {Lewis}}]{fortney2010}
{Fortney}, J.~J., {Shabram}, M., {Showman}, A.~P., {et~al.} 2010, \apj, 709,
  1396

\bibitem[{{Goodman}(2009)}]{goodman2009}
{Goodman}, J. 2009, \apj, 693, 1645

\bibitem[{{Grillmair} {et~al.}(2008){Grillmair}, {Burrows}, {Charbonneau},
  {Armus}, {Stauffer}, {Meadows}, {van Cleve}, {von Braun}, \&
  {Levine}}]{grillmair2008}
{Grillmair}, C.~J., {Burrows}, A., {Charbonneau}, D., {et~al.} 2008, \nat, 456,
  767

\bibitem[{{Harrington} {et~al.}(2006){Harrington}, {Hansen}, {Luszcz},
  {Seager}, {Deming}, {Menou}, {Cho}, \& {Richardson}}]{harrington2006}
{Harrington}, J., {Hansen}, B.~M., {Luszcz}, S.~H., {et~al.} 2006, Science,
  314, 623

\bibitem[{{Heng} {et~al.}(2011){Heng}, {Frierson}, \& {Phillipps}}]{heng2011}
{Heng}, K., {Frierson}, D.~M.~W., \& {Phillipps}, P.~J. 2011, ArXiv e-prints

\bibitem[{{Karoly} {et~al.}(1998){Karoly}, {Vincentt}, \&
  {Schrage}}]{karoly1998}
{Karoly}, D.~J., {Vincentt}, D.~G., \& {Schrage}, J.~M., eds. 1998, General
  Circulation. Meterology of the Southern Hemisphere, Vol.~27

\bibitem[{{Kley} \& {Hensler}(1987)}]{kley1987}
{Kley}, W. \& {Hensler}, G. 1987, \aap, 172, 124

\bibitem[{{Knutson} {et~al.}(2008){Knutson}, {Charbonneau}, {Allen}, {Burrows},
  \& {Megeath}}]{knutson2008}
{Knutson}, H.~A., {Charbonneau}, D., {Allen}, L.~E., {Burrows}, A., \&
  {Megeath}, S.~T. 2008, \apj, 673, 526

\bibitem[{{Knutson} {et~al.}(2007){Knutson}, {Charbonneau}, {Allen}, {Fortney},
  {Agol}, {Cowan}, {Showman}, {Cooper}, \& {Megeath}}]{knutson2007}
{Knutson}, H.~A., {Charbonneau}, D., {Allen}, L.~E., {et~al.} 2007, \nat, 447,
  183

\bibitem[{{Knutson} {et~al.}(2009){Knutson}, {Charbonneau}, {Cowan}, {Fortney},
  {Showman}, {Agol}, {Henry}, {Everett}, \& {Allen}}]{knutson2009}
{Knutson}, H.~A., {Charbonneau}, D., {Cowan}, N.~B., {et~al.} 2009, \apj, 690,
  822

\bibitem[{{Langton} \& {Laughlin}(2007)}]{langton2007}
{Langton}, J. \& {Laughlin}, G. 2007, \apjl, 657, L113

\bibitem[{{Langton} \& {Laughlin}(2008)}]{langton2008}
---. 2008, \apj, 674, 1106

\bibitem[{{Lecavelier Des Etangs} {et~al.}(2008){Lecavelier Des Etangs},
  {Vidal-Madjar}, {D{\'e}sert}, \& {Sing}}]{desetangs2008}
{Lecavelier Des Etangs}, A., {Vidal-Madjar}, A., {D{\'e}sert}, J., \& {Sing},
  D. 2008, \aap, 485, 865

\bibitem[{{Leconte} {et~al.}(2011){Leconte}, {Lai}, \&
  {Chabrier}}]{leconte2011}
{Leconte}, J., {Lai}, D., \& {Chabrier}, G. 2011, \aap, 528, A41+

\bibitem[{{Levermore} \& {Pomraning}(1981)}]{levermore1981}
{Levermore}, C.~D. \& {Pomraning}, G.~C. 1981, \apj, 248, 321

\bibitem[{{Li} \& {Goodman}(2010)}]{li2010_2}
{Li}, J. \& {Goodman}, J. 2010, \apj, 725, 1146

\bibitem[{{Mandel} \& {Agol}(2002)}]{mandel2002}
{Mandel}, K. \& {Agol}, E. 2002, \apjl, 580, L171

\bibitem[{{Menou} \& {Rauscher}(2009)}]{menou2009}
{Menou}, K. \& {Rauscher}, E. 2009, \apj, 700, 887

\bibitem[{{Perna} {et~al.}(2010){Perna}, {Menou}, \& {Rauscher}}]{perna2010}
{Perna}, R., {Menou}, K., \& {Rauscher}, E. 2010, ArXiv e-prints

\bibitem[{{Rauscher} \& {Menou}(2010)}]{rauscher2010}
{Rauscher}, E. \& {Menou}, K. 2010, \apj, 714, 1334

\bibitem[{{Rauscher} {et~al.}(2008){Rauscher}, {Menou}, {Cho}, {Seager}, \&
  {Hansen}}]{rauscher2008}
{Rauscher}, E., {Menou}, K., {Cho}, J.~Y.-K., {Seager}, S., \& {Hansen},
  B.~M.~S. 2008, \apj, 681, 1646

\bibitem[{{Seager} \& {Hui}(2002)}]{seager2002}
{Seager}, S. \& {Hui}, L. 2002, \apj, 574, 1004

\bibitem[{{Seager} \& {Sasselov}(2000)}]{seager2000}
{Seager}, S. \& {Sasselov}, D.~D. 2000, \apj, 537, 916

\bibitem[{{Shabram} {et~al.}(2011){Shabram}, {Fortney}, {Greene}, \&
  {Freedman}}]{shabram2011}
{Shabram}, M., {Fortney}, J.~J., {Greene}, T.~P., \& {Freedman}, R.~S. 2011,
  \apj, 727, 65

\bibitem[{{Sharp} \& {Burrows}(2007)}]{sharp2007}
{Sharp}, C.~M. \& {Burrows}, A. 2007, \apjs, 168, 140

\bibitem[{{Showman} {et~al.}(2008){Showman}, {Cooper}, {Fortney}, \&
  {Marley}}]{showman2008_2}
{Showman}, A.~P., {Cooper}, C.~S., {Fortney}, J.~J., \& {Marley}, M.~S. 2008,
  \apj, 682, 559

\bibitem[{{Showman} {et~al.}(2009){Showman}, {Fortney}, {Lian}, {Marley},
  {Freedman}, {Knutson}, \& {Charbonneau}}]{showman2009}
{Showman}, A.~P., {Fortney}, J.~J., {Lian}, Y., {et~al.} 2009, \apj, 699, 564

\bibitem[{{Showman} \& {Guillot}(2002)}]{showman2002}
{Showman}, A.~P. \& {Guillot}, T. 2002, \aap, 385, 166

\bibitem[{{Showman} \& {Polvani}(2011)}]{showman2011}
{Showman}, A.~P. \& {Polvani}, L.~M. 2011, ArXiv e-prints

\bibitem[{{Snellen} {et~al.}(2010){Snellen}, {de Kok}, {de Mooij}, \&
  {Albrecht}}]{snellen2010}
{Snellen}, I.~A.~G., {de Kok}, R.~J., {de Mooij}, E.~J.~W., \& {Albrecht}, S.
  2010, \nat, 465, 1049

\end{thebibliography}

\end{document}